\documentclass[10pt]{article}
\usepackage{latexsym}
\usepackage{amssymb}
\usepackage{amsmath}
\usepackage{amscd}
\usepackage{amsthm}
\usepackage[left=2cm,top=2.5cm,right=2.5cm,bottom=1.5cm]{geometry}

\usepackage[dvips]{graphicx}
\usepackage{hyperref}

\begin{document}
\begin{center}
\Large{\bf{Transit universe with time varying $G$ and decaying $\Lambda$}}
\vspace{10mm}

\normalsize{Anil Kumar Yadav $\footnote{corresponding author}$, Amit Sharma$^2$}\\ 
\vspace{4mm} 
\normalsize{$^1$Department of Physics, Anand Engineering
College, Keetham, Agra-282 007, India} \\
\vspace{2mm}
\normalsize{$^{1}$E-mail: abanilyadav@yahoo.co.in}\\
\vspace{4mm}
\normalsize{$^{2}$Department of Mathematics, Anand Engineering College, Keetham, Agra-282 007, India} 
\end{center}
\begin{abstract} 
We present the model of transit universe from early deceleration phase to current 
acceleration phase under the framework of general relativity in presence of gravitational 
coupling $G(t)$ and cosmological terms $\Lambda(t)$. The Einstein's field equations have 
been solved by considering time dependent deceleration parameter (DP) which renders 
the scale factor $a = (t^{n}e^{kt})^{\frac{1}{m}}$ where $m$, $n$ and $k$ are positive constants.
The cosmological term $(\Lambda(t))$ is found to be positive and a decreasing function of 
time which supports the result obtained from supernovae Ia observations. The geometrical 
and kinematical features of the model are examined in detail.\\ 
\end{abstract}


\section{Introduction}
The most striking discovery of the modern physics is that the current Universe 
is not only expanding but also accelerating. The late time accelerated expansion of 
Universe has been confirmed by SN Ia observations (Riess et al. 1998; Perlmutter et al. 1999). 
Observations also suggest that the had been a transition of Universe from the 
earlier deceleration phase to current acceleration phase (Caldwell et al. 2004). 
The recent measurements of the CMB anisotropy and the observations from type Ia supernovae demand
a significant and positive cosmological constant (Perlmutter 1997, Fujii 2000). 
Also observations of gravitational lensing indicate the presence of a non-zero $\Lambda$.\\

The cosmological term $\Lambda$ and the gravitational coupling G are assumed to be constants 
in the Einstein's theory of general relativity. 
However, alternative ideas about the variability of these parameters have been
started long ago. The idea of variable G was first introduced by Dirac (1937), though 
Dirac's arguments were based cosmological considerations not directly directly 
concerned with Mach's principle. Later on Brans and Dike (1961) formulated the scalar-
tensor theory of gravitation which is based on the coupling between an adequate 
tensor field and a scalar field $\phi$, having the dimension of $G^{-1}$. Motivated by dimensional grounds with quantum 
cosmology, Chen and wu (1990) have consider the variation of cosmological term as $\Lambda\propto R^{-2}$. 
However, a number of authors have argued in favour of the dependence $\Lambda \propto t^{-2}$. 
Later on, Arbab (2003) has investigated cosmic acceleration with positive cosmological constant. 
A positive cosmological constant helps overcome the age problem, connected on the one side 
with the high estimates of the Hubble parameters and with the age of the globular 
clusters on the other. Further, it seems that in order to retain the cold dark matter 
theory in the spatially flat Universe most of the critical density should be provided 
by a positive cosmological constant (Efstathiou el al 1990, Kofman et al 1993). Observational 
data indicates that the cosmological constant, if nonzero, is smaller than $10^{-55}$ cm$^{-2}$. 
However, since everything that contributes to the vacuum energy acts as a 
cosmological constant it can not just be dropped without serious considerations. Moreover 
particle physics expectations for $\Lambda$ exceeds its present value by the factor of 
order $10^{120}$ i. e. in a sharp contrast to observations. To explain this apparent 
discrepancy the point of view has been adapted which allow the $\Lambda$-term to vary with time 
(Salim and Waga 1993, Matyjasel 1995). The idea of that during the evolution of universe the 
energy density. The idea of that during the evolution of Universe the energy density 
of the vacuum decays into the particles thus leading to the decrease of cosmological 
constant. As a result one has the creation of particles although the typical rate 
of the creation is very small.\\

An anisotropic Bianchi type V cosmological model plays a significant role 
in understanding the phenomenon like formation of galaxies during its early stage of 
evolution. The choice of anisotropic cosmological models permit one to obtain 
more general cosmological model in comparison to FRW model. Theoretical arguments 
and recent observations of CMBR support the existence of an anisotropic phase 
that approaches an isotropic one. Therefore, it makes sense to consider the models of 
Universe with anisotropic background in presence of gravitational coupling $G$ and 
cosmological term $\Lambda$. Among different anisotropic cosmological models Bianchi type V Universe is 
natural generalization of the open FRW model. Lorentz (1981, 1985) investigated 
tilted Bianchi type V cosmological model with matter and electromagnetic 
field in higher dimensions. A large number of authors have studied Bianchi type V 
cosmological model in different contexts (Beesham 1986; Benerjee and Sanyal 1988; 
Nayak and Sahoo 1989, 1996; Coley 1990; Singh and Singh 1991; Coley and Dunn 1992; Pradhan and Rai 2004). 
Singh and Chaubey (2006) have considered 
a Bianchi type V universe initially for self consistent system of gravitational 
field with a binary mixture of perfect fluid and dark energy given by a cosmological 
constant. Further they have studied the evolution of a homogeneous anisotropic Universe 
filled with viscous fluid, in the presence of cosmological constant $\Lambda$ (2007). 
Singh and Kale (2009) and recently Yadav et al (2012) have discussed anisotropic 
bulk viscous cosmological models with variable $G$ and $\Lambda$.\\

In this paper, we present the model of transit Universe with $G(t)$ and $\Lambda(t)$. 
To study the transit behaviour of Universe, we assume the scale factor as an increasing 
function of time which generates a time dependent deceleration parameter (DP). The paper 
is organized as follows. In section 2, the model and field equations have been presented. 
Section 3 deals with the scale factors and cosmological parameters. Finally conclusions 
are presented in Section 5.\\

\section{Model and Field equations}
We consider the space-time admitting Bianchi type-V group of motion in the form
\begin{equation}
 \label{eq1}
ds^{2}=-dt^{2}+A^{2}dx^{2}+e^{2\alpha x}\left(B^{2}dy^{2}+C^{2}dz^{2}\right)
\end{equation}
where $A(t)$, $B(t)$ and $C(t)$ are the scale factors in x, y and z directions and 
$\alpha$ is a constant.\\
The average scale factor $(a)$ and spatial volume of Bianchi-type V metric are given by
\begin{equation}
 \label{eq2}
a=(ABC)^{\frac{1}{3}}
\end{equation}
\begin{equation}
\label{eq3}
V = a^{3} = ABC 
\end{equation}
The generalized mean Hubble's parameter $(H)$ is given by 
\begin{equation}
 \label{eq4}
H=\frac{\dot{a}}{a}=\frac{1}{3}\left(H_{x}+H_{y}+H_{z}\right)
\end{equation}
where $H_{x}=\frac{\dot{A}}{A}$, $H_{y}=\frac{\dot{B}}{B}$ and $H_{z}=\frac{\dot{C}}{C}$ are the 
directional Hubble's parameters. An over dot denotes 
differentiation with respect to cosmic time t.\\
Since metric (\ref{eq1}) is completely characterized by average scale factor therefore let 
us consider that the average scale factor is increasing function of time as following
\begin{equation}
\label{eq5}
a=(t^{n}e^{kt})^{\frac{1}{m}} 
\end{equation}
where $k\geq 0$, $m > 0$ and $n \geq 0$ are constant. 
It is important to note here that the ansatz for scale factor generalized the one proposed by 
Yadav (2012a, 2012b) and Pradhan $\&$ Amirhashchi (2011). Yadav (2012a, 2012b) considered 
string and bulk viscous fluid as source of matter to describe the transit behaviour of universe whereas 
Pradhan $\&$ Amirhashchi (2011) studied dark energy model with variable EOS parameter. In this paper, we 
considered cosmic fluid filled with $G(t)$ and $\Lambda(t)$ as source of matter to describe 
the transition of universe from early decelerating phase to 
current accelerating phase.\\

The value of DP (q) for model (\ref{eq1})is found to be
\begin{equation}
\label{eq6}
q=-\frac{\ddot{a}a}{\dot{a}^{2}} = -1+\frac{mn}{(n+kt)^{2}}
\end{equation}
Equation (\ref{eq6}) clearly indicates the time varying nature of DP (q). 
Amendola (2003) and Riess et al (2001) found that the expansion of 
Universe is accelerating at present epoch but it was decelerating in past and 
the transition redshift from decelerated to present accelerated expansion is about 0.5.  
It is however possible to have $n =0$ in equation (\ref{eq5}) for which we would have 
inflationary universe. The sign of $q$ indicates whether the model inflates or not. A 
positive sign of $q$ i. e. $t \leq \frac{1}{k}\left[\sqrt{mn} - n\right]$ 
corresponds to standard decelerating model whereas 
the negative sign $-1\leq q <0$ indicates the inflation.\\

The Einstein's field equations read as
\begin{equation}
 \label{eq7}
R^{i}_{j}-\frac{1}{2}g^{i}_{j}R-\Lambda g^{i}_{j} = -8\pi G T^{i}_{j}
\end{equation}
where $T^{i}_{j}$ is the energy momentum tensor and it is given by
\begin{equation}
 \label{eq8}
T^{i}_{j} = (\rho + p)v^{i}v_{j}-pg^{i}_{j}
\end{equation}
where $\rho$ is the energy density, $p$ is the isotropic pressure of 
    the cosmic fluid, and $v^{i}$ is the fluid four velocity vector. In 
co-moving system of co-ordinates, we have $v^{i} = (1,0,0,0)$.\\

The Einstein's field equation (\ref{eq7}) for the line-element (\ref{eq1}) 
leads to the following system of equations
\begin{equation}
 \label{eq9}
\frac{\ddot{B}}{B}+\frac{\ddot{C}}{C}+\frac{\dot{B}\dot{C}}{BC}-\frac{\alpha^{2}}{A^{2}} = -8\pi G \gamma\rho + \Lambda
\end{equation}
\begin{equation}
 \label{eq10}
\frac{\ddot{A}}{A}+\frac{\ddot{C}}{C}+\frac{\dot{A}\dot{C}}{AC}-\frac{\alpha^{2}}{A^{2}} = -8\pi G \gamma\rho + \Lambda
\end{equation}
\begin{equation}
 \label{eq11}
\frac{\ddot{A}}{A}+\frac{\ddot{B}}{B}+\frac{\dot{A}\dot{B}}{AB}-\frac{\alpha^{2}}{A^{2}} = -8\pi G \gamma\rho + \Lambda
\end{equation}
\begin{equation}
 \label{eq12}
\frac{\dot{A}\dot{B}}{AB}+\frac{\dot{A}\dot{C}}{AC}+\frac{\dot{B}\dot{C}}{BC}-\frac{3\alpha^{2}}{A^{2}} = \rho + \Lambda
\end{equation}
\begin{equation}
 \label{eq13}
\frac{2\dot{A}}{A}-\frac{\dot{B}}{B}-\frac{\dot{C}}{C} = 0
\end{equation}
Here, we have assumed, as usual, an equation of state $p = \gamma \rho$, 
where $0\leq \gamma \leq 1$ is constant.\\

The shear scalar $(\sigma)$ is obtained as
\begin{equation}
 \label{eq14}
\sigma^{2} = \frac{1}{3}\left[\frac{\dot{A}^{2}}{A^{2}}+\frac{\dot{B}^{2}}{B^{2}} \frac{\dot{C}^{2}}{C}-
\left(\frac{\dot{A}\dot{B}}{AB}+\frac{\dot{B}\dot{C}}{BC}+\frac{\dot{A}\dot{C}}{AC}\right)\right]
\end{equation}
Equations (\ref{eq4}), (\ref{eq12}) and (\ref{eq14}) allow to write the analogue 
of Friedmann equation as
\begin{equation}
 \label{eq15}
3H^{2} = 8\pi G \rho + \sigma^{2} + \Lambda + \frac{3\alpha^{2}}{A^{2}}
\end{equation}
Here, we obtain that same equations as in case of constant $G$ and $\Lambda$; therefore 
the varying characters of $G$ and $\Lambda$ do not affect the equations. Finally the generalized 
conservation equation can be obtained by using equations (\ref{eq9}) - 
(\ref{eq11}) in the differentiated form of equation (\ref{eq12}) and can be written, in the form
\begin{equation}
 \label{eq16}
8\pi G\left[\dot{\rho}+3(1+\gamma)\rho H\right]+8\pi \rho \dot{G} + \dot{\Lambda} = 0
\end{equation}
We assume that the conservation of 
energy momentum tensor of matter holds $(T^{ij}_{;\;j} = 0)$ leading to 
\begin{equation}
 \label{eq17}
\dot{\rho} + 3(1+\gamma)\rho H = 0
\end{equation}
leaving $G$ and $\Lambda$ as some kind of coupled fields
\begin{equation}
 \label{eq18}
8\pi \rho \dot{G} + \Lambda = 0
\end{equation}
\section{The scale factors and cosmological parameters}
Integrating equation (\ref{eq13}) and absorbing the constant of 
integration into $B$ or $C$, we obtain
\begin{equation}
\label{eq19}
 A^{2} = BC
\end{equation}
From equations (\ref{eq9}) - (\ref{eq11}) and taking second integral of each, 
we obtain the following three relations, respectively,
\begin{equation}
\label{eq20}
\frac{A}{B} = b_{1}\;exp\left(x_{1}\int a^{-3}dt\right)
\end{equation}
\begin{equation}
\label{eq21}
\frac{A}{C} = b_{2}\;exp\left(x_{2}\int a^{-3}dt\right)
\end{equation} 
\begin{equation}
\label{eq22}
\frac{B}{C} = b_{3}\;exp\left(x_{3}\int a^{-3}dt\right)
\end{equation}
where $b_{1}$, $b_{2}$, $b_{3}$, $x_{1}$, $x_{2}$ and $x_{3}$ are constant of integrations.\\

From equations (\ref{eq19}) - (\ref{eq22}) and (\ref{eq5}), the metric functions can be explicitly 
written as
\begin{equation}
\label{eq23}
A(t) = (t^{n}e^{kt})^{\frac{1}{m}}
\end{equation}
\begin{equation}
\label{eq24}
B(t)=d\;(t^{n}e^{kt})^{\frac{1}{m}}\;exp\left(\ell\int(t^{n}e^{kt})^{-\frac{3}{m}}dt\right)
\end{equation}
\begin{equation}
\label{eq25}
C(t)=d^{-1}\;(t^{n}e^{kt})^{\frac{1}{m}}\;exp\left(-\ell\int(t^{n}e^{kt})^{-\frac{3}{m}}dt\right)
\end{equation}
where $d = (b_{2}b_{3})^{\frac{1}{3}}$, $\ell = \frac{x_{2}+x_{3}}{3}$ with $b_{2} = b_{1}^{-1}$ and $x_{2} = - x_{1}$.\\\

\begin{figure}
\begin{center}
\includegraphics[width=4in]{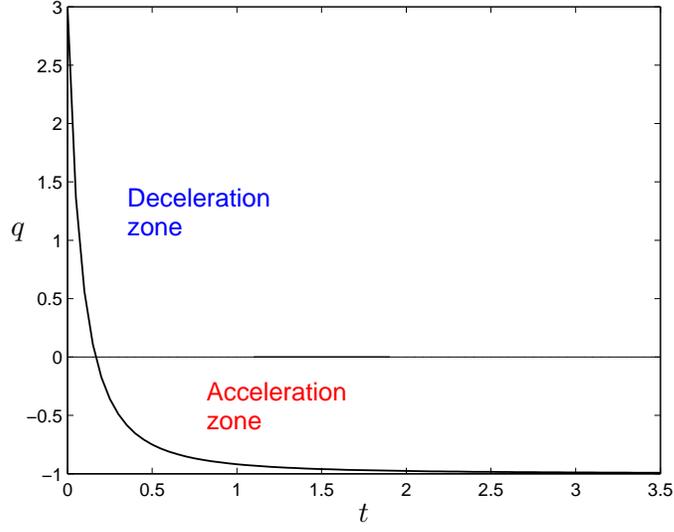} 
\caption{The plot of DP $(q)$ vs. time (t) with $m = 1$, $n = 0.25$, $k = 1.5$.}
\label{fg:anil40F1.eps}
\end{center}
\end{figure}

Integrating equation (\ref{eq17}), we obtain
\begin{equation}
 \label{eq26}
\rho = \rho_{0}(t^{n}e^{kt})^{-\frac{3(1+\gamma)}{m}}
\end{equation}
where $\rho_{0}$ is the positive constant of integration.\\

The physical parameters such as directional Hubble parameters $(H_{x}, H_{y}, H_{z})$, 
average Hubble parameter (H), shear scalar $(\sigma)$, expansion scalar $(\theta)$ and 
spatial volume $(V)$ are given by
\begin{equation}
 \label{eq27}
H_{x}=\frac{1}{m}\left(\frac{n}{t}+k\right)
\end{equation}
\begin{equation}
 \label{eq28}
H_{y}=\frac{1}{m}\left(\frac{n}{t}+k\right)+\ell\left(t^{n}e^{kt}\right)^{-\frac{3}{m}}
\end{equation}
\begin{equation}
 \label{eq29}
H_{y}=\frac{1}{m}\left(\frac{n}{t}+k\right)-\ell\left(t^{n}e^{kt}\right)^{-\frac{3}{m}}
\end{equation}
\begin{equation}
 \label{eq30}
H=\frac{1}{m}\left(\frac{n}{t}+k\right)
\end{equation}
\begin{equation}
 \label{eq31}
\sigma^{2}= \ell (t^{n}e^{kt})^{-\frac{6}{m}}
\end{equation}
\begin{equation}
 \label{eq32}
\theta=\frac{3}{m}\left(\frac{n}{t}+k\right)
\end{equation}
Equations (\ref{eq31}) and (\ref{eq32}) lead to
\begin{equation}
 \label{eq33}
\frac{\sigma}{\theta}=\frac{\sqrt{\ell}m}{3}(t^{n}e^{kt})^{-\frac{3}{m}}(\frac{n}{t}+k)^{-1}
\end{equation}
From equation (\ref{eq33}), we observe that $lim_{t\rightarrow \infty}\left(\frac{\sigma}{\theta}\right) = 0$. Thus 
the derived model approaches to isotropy at present epoch. Figure 1 shows the dynamics of 
DP from early deceleration phase to recent acceleration phase whereas Figure 2 ensures 
that $H_{x}$, $H_{y}$ and $H_{z}$ evolve with equal rate at late times therefore  
the Universe achieves isotropy at present epoch.\\

It is observed that the scale factors $A(t)$, $B(t)$ and $C(t)$ along the spatial 
directions x, y and z respectively, vanish at $t=0$. Thus the model has a point 
type singularity at $t = 0$. We obtain $q = -1$ and $\frac{dH}{dt} = 0$ as $t\rightarrow \infty$. The 
model under consideration has time dependent DP and evolves to isotropy as $t\rightarrow \infty$, with 
$\Lambda\rightarrow 0$. Thus for large time, the model approaches the flat FLRW model which is very 
encouraging. It may be noted that through the current observations of SN Ia and CMB 
favour accelerating models $q < 0$, but they do not altogether rule out the decelerating ones which 
are also consistent with these observations. It is possible to fit the model with zero $\Lambda$, considering 
the extinction of light by the metallic dust ejected 
from the supernovae explosions (Viswakarma 2003).\\

\begin{figure}
\begin{center}
\includegraphics[width=4in]{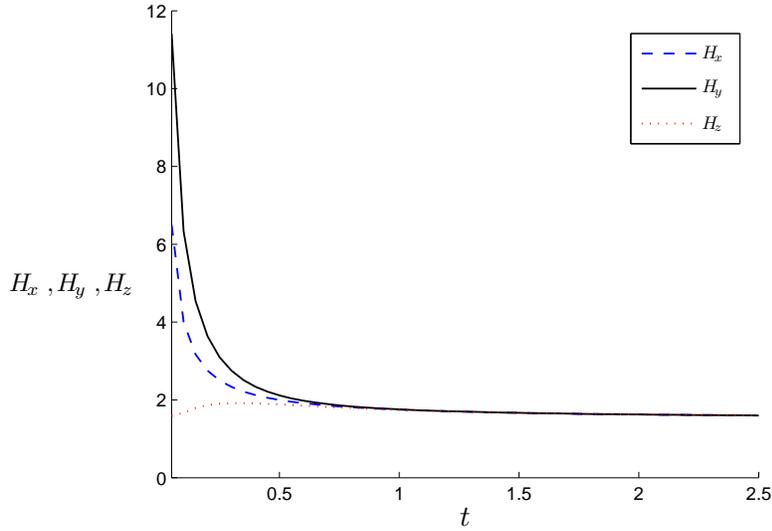} 
\caption{The plot of directional Hubble parameters vs. time with $m = 1$, $n = 0.25$, $k = 1.5$ and 
$\ell = 0.65$.}
\label{fg:anil40F2.eps}
\end{center}
\end{figure}
\begin{figure}
\begin{center}
\includegraphics[width=4in]{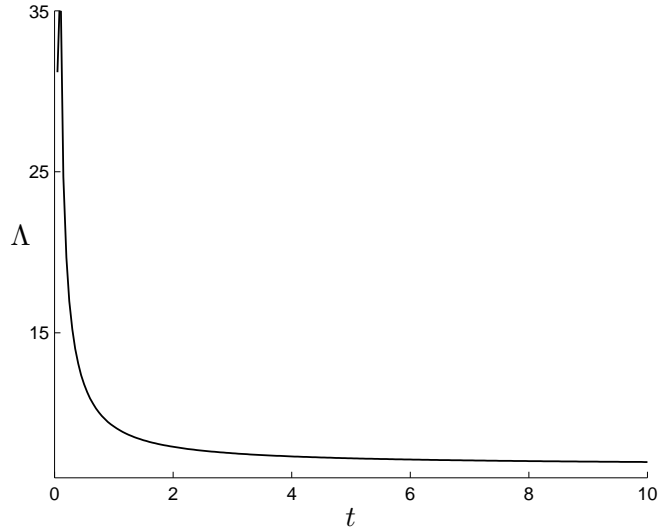} 
\caption{Cosmological constant $(\Lambda)$ vs. time (t) with $m = 1$, $n =0.25$, $k = 1.5$, $\gamma = 0.33$, 
$\alpha = 0.15$ and $\rho_{0} = 1$.}
\label{fg:anil40F3.eps}
\end{center}
\end{figure}
The cosmological constant $(\Lambda)$ and Gravitational constant $(G)$ are 
found to be
\begin{equation}
 \label{eq34}
\Lambda = \frac{3}{m^{2}}\left(\frac{n}{t}+k\right)^{2}-\ell^{2}\left(t^{n}e^{kt}\right)^{-\frac{6}{m}}-
3\alpha^{2}\left(t^{n}e^{kt}\right)^{-\frac{2}{m}} - \rho_{0}(t^{n}e^{kt})^{-\frac{3(1+\gamma)}{m}}
\end{equation}
\begin{equation}
 \label{eq35}
G = \frac{m}{24\pi(1+\gamma)}\left(t^{n}e^{kt}\right)^{\frac{3(1+\gamma)}{m}}
\left[\frac{6n}{m^{2}t^{2}}-\frac{6(\ell^{2}+\alpha^{2})}{m}t^{n}e^{kt}\right]
\end{equation}

From equation (\ref{eq35}), it is observed that cosmological constant $(\Lambda)$ 
is decreasing function of time. This behaviour is clearly shown in Fig. 3. 
Recent cosmological observations suggest the existence of positive cosmological 
constant $(\Lambda)$ with the magnitude $\Lambda\left(\frac{G\hbar}{c^{3}}\right)\approx 10^{-123}$. 
These observations on magnitude and ref-shift of type Ia supernova suggest that our 
Universe may be an accelerating one with induced cosmological density through the 
cosmological $\Lambda$-term. Thus the model presented in this paper 
is consistent with the results of recent observations.\\

We can express, equations (\ref{eq9})$-$(\ref{eq12}), in terms of $H$, $q$ and $\sigma$ as
\begin{equation}
 \label{eq36}
8\pi G\gamma\rho-\Lambda = (2q-1)H^{2}-\sigma^{2}+\frac{\alpha^{2}}{a^{2}}
\end{equation}
\begin{equation}
 \label{eq37}
8\pi G\gamma\rho-\Lambda = 3H^{2}-\sigma^{2}-\frac{3\alpha^2}{a^{2}}
\end{equation}
Equations (\ref{eq36}) and (\ref{eq37}) leads to
\begin{equation}
 \label{eq38}
\frac{\ddot{a}}{a} = \frac{\Lambda}{3}-\frac{2\sigma^{2}}{3}-\frac{1}{6}8\pi G (\rho + 3p)
\end{equation}
From equation (\ref{eq38}), it is shown that for $\rho + 3p = 0$, only $\Lambda$-terms contributes 
the acceleration which seems to 
relate $\Lambda$ with dark energy. The same is predicted by 
supernovae legacy survey (SNLS) observations.\\

\begin{figure}
\begin{center}
\includegraphics[width=4in]{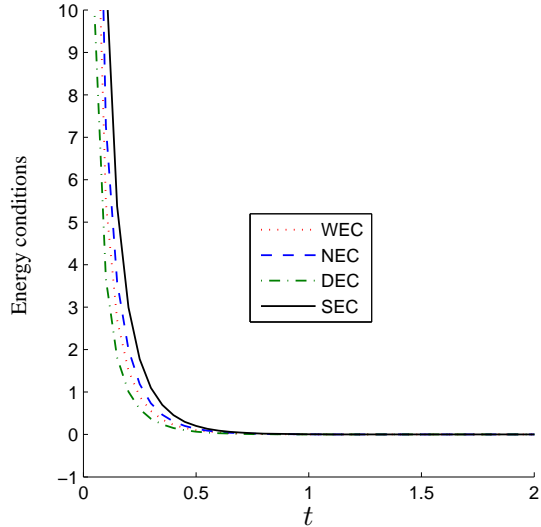} 
\caption{The left hand side of energy conditions vs. time with $m = 1$, $n =0.25$, $k = 1.5$, $\gamma = 0.33$ 
and $\rho_{0} = 1$.}
\label{fg:anil40F4.eps}
\end{center}
\end{figure}

\section{Results and Summary}
In this paper, we have presented the model of transit Universe with 
gravitational coupling $G(t)$ and cosmological term $\Lambda(t)$ in the framework of general relativity.
The spatial scale factors 
and volume scalar of derived model vanish at $t = 0$. 
The energy density and pressure are infinite at this initial epoch. 
As $t \rightarrow \infty$, the scale factor diverge and $\rho$ tends to zero. The shear scalar $(\sigma)$ 
are very large at initial moment but decrease with cosmic time and vanish 
at $t \rightarrow \infty$. The model shows isotropic state in later time of its evolution. For $n\neq 0$, 
all matter and radiation is concentrated 
at the big bang epoch and the model has a point 
type singularity at the initial moment. For $n = 0$, the universe has 
non singular origin which seems reasonable to project the dynamics of future Universe. 
In the derived model, $lim_{t\rightarrow 0}\frac{\rho}{\theta^{2}}$ turns out to be constant. 
Thus matter is dynamically negligible near the origin and the model approaches homogeneity.\\

The cosmological constant $(\Lambda)$ is found to be decreasing 
function of time and it approaches to small positive value at late time. 
A positive value of $\Lambda$ corresponds to negative effective mass density (repulsion). 
Hence we expect that in the Universe with the positive value of $\Lambda$, the expansion will 
tends to accelerate. Thus the derived model predicts accelerating universe at present epoch.\\
 
The age of the Universe, in the derived model is given by 
$$T_{0} = \frac{n}{m}H_{0}^{-1}-k$$
which differs from the present estimate i. e. $T_{0} = H_{0}^{-1}\approx 14$ Gyr. 
But if we take $n = m$ and $k = 0$, the model is in good agreement with present age 
of Universe. The energy conditions are satisfied (Figure 4) which turns to imply that 
model presented in this paper is physically realistic. Finally a physically viable 
model of transit universe from early deceleration phase to current acceleration phase 
with singular origin has been obtained.\\


\end{document}